\documentclass[aps,prev,twocolumn,preprintnumbers,floatfix,nofootinbib]{revtex4-1}
\pdfoutput=1
\usepackage[english]{babel}
\usepackage{graphicx,color}
\usepackage{bm}
\usepackage{times}
\usepackage{slashed}
\usepackage{multirow}
\usepackage[usenames,dvipsnames,svgnames,table]{xcolor}
\usepackage{slashed}
\usepackage{amsmath}
\usepackage{amsthm}
\usepackage{subfigure}
\usepackage{hyperref}
\usepackage[dvipsnames]{xcolor}
\definecolor{bluencs}{rgb}{0.0, 0.53, 0.74}
\definecolor{darkcyan}{rgb}{0.0, 0.55, 0.55}
\definecolor{hanblue}{rgb}{0.27, 0.42, 0.81}
\hypersetup{
     linktocpage=true,
     setpagesize=true,
     urlcolor=blue,
     citecolor=blue,
     linkcolor=blue, 
     menucolor=cyan,
     colorlinks=true, 
     filecolor=magenta,
     citebordercolor=blue,
     pdftitle={Superposition of CP-Even and CP-Odd Higgs Resonances:Explaining the 95 GeV Excesses within a Two-Higgs Doublet Model},
     pdfsubject={Latex},
     pdfauthor={Benbrik, Boukidi, Moretti},
    pdfkeywords={BSM},
    unicode=true
}
\usepackage[capitalize]{cleveref}

\newcommand{\be}{\begin{equation}}
\newcommand{\ee}{\end{equation}}
\newcommand{\bea}{\begin{eqnarray}}
\newcommand{\eea}{\end{eqnarray}}

\usepackage{xfrac}
 
\usepackage{float}
\usepackage{bbding,pifont}
\definecolor{brilliantrose}{rgb}{1.0, 0.33, 0.64}
\definecolor{lawngreen}{rgb}{0.49, 0.99, 0.0}
\definecolor{magenta}{rgb}{1.0, 0.0, 1.0}
\makeatletter
\let\old@float\@float
\def\@float{\let\centering\relax\old@float}
\makeatother
\begin{document}

\title{Superposition of CP-Even and CP-Odd Higgs Resonances:\\[0.15cm]
	Explaining the 95 GeV Excesses within a Two-Higgs Doublet Model}

\author{Rachid Benbrik$^{a}$}
\email{r.benbrik@uca.ac.ma}

\author{Mohammed Boukidi$^{a}$}
\email{mohammed.boukidi@ced.uca.ma}

\author{Stefano Moretti$^{b,c}$}
\email{s.moretti@soton.ac.uk}\email{stefano.moretti@physics.uu.se}

\affiliation{$^a$Polydisciplinary Faculty, Laboratory of Fundamental and Applied Physics, Cadi Ayyad University, Sidi Bouzid, B.P. 4162, Safi, Morocco.}
\affiliation{$^b$School of Physics and Astronomy, University of Southampton,\\ Southampton, SO17 1BJ, United Kingdom.}
\affiliation{$^c$Department of Physics and Astronomy, Uppsala University,\\ Box 516, SE-751 20 Uppsala, Sweden.}

\begin{abstract}
We propose an explanation for the observed excesses around 95 GeV in the di-photon and di-tau invariant mass distributions, as reported by the CMS collaboration at the Large Hadron Collider (LHC). These findings are complemented by a long-standing discrepancy in the $b\bar{b}$ invariant mass at the Large Electron-Positron (LEP) Collider. Additionally, the ATLAS collaboration has reported a corroborative excess in the di-photon final state within the same mass range, albeit with slightly lower significance.
Our approach involves the superposition of CP-even and CP-odd Higgs bosons within the Type-III Two-Higgs Doublet Model (2HDM) to simultaneously explain these excesses at 1$\sigma$ Confidence Level (C.L.), while remaining consistent with current theoretical and experimental constraints.
\end{abstract}

\maketitle

\section{Introduction}
\noindent 
In the last decade, following the Higgs boson's discovery at the Large Hadron Collider (LHC) in 2012~\cite{ATLAS:2012yve,CMS:2012qbp}, the scientific community has made significant strides towards the precise characterization of its properties. These efforts have affirmed the SM (SM) predictions with accuracies rarely exceeding 10\%. Despite these achievements, the quest for physics Beyond the SM  (BSM) persists, encouraged by the precision of current Higgs physics at the LHC. This has opened the door to exploring additional Higgs states beyond the SM-like one, ranging in mass from a few GeV to the TeV scale. Extended Higgs sectors, as anticipated in various BSM scenarios including Supersymmetric models \cite{Moretti:2019ulc} and 2-Higgs Doublet Models (2HDMs)~\cite{Gunion:1992hs,Branco:2011iw}, suggest the presence of both light and heavy non-standard Higgs bosons. These predictions have spurred searches for these (pseudo)scalar states across lepton and hadron colliders.

The 2HDM is a particularly well-studied framework within BSM theories, extending the Standard Model (SM) Higgs sector by an additional Higgs doublet. Its general version allows non-diagonal Yukawa couplings, potentially leading to Flavor Changing Neutral Currents (FCNCs) at tree level, contrary to experimental evidence. To circumvent this issue, a $Z_2$ symmetry is typically imposed to define the coupling structure of the two Higgs doublets to SM fermions. This classification includes the so-called Type-I, Type-II, lepton-specific and flipped scenarios \cite{Branco:2011iw}, alongside the 2HDM Type-III, which allows direct couplings of both doublets to all SM fermions. Its Yukawa structure is then refined by both theoretical consistency requirements and experimental measurements of Higgs masses and couplings.

During ongoing searches for a low-mass Higgs boson, the CMS collaboration reported an excess near 95 GeV in di-photon event invariant masses in 2018 \cite{CMS:2018cyk}. In March 2023, CMS confirmed this excess with a local significance of 2.9$\sigma$ at $m_{\gamma\gamma}=95.4$ GeV, employing advanced analyses on data from Run 2's first three years \cite{CMS:2023yay}. Similarly, ATLAS observed an excess at 95 GeV with a local significance of 1.7$\sigma$, aligning with CMS' findings and showcasing enhanced sensitivity over previous analyses \cite{Arcangeletti, ATLAS:2023jzc}.

Moreover, CMS has reported an excess in the search for a light neutral (pseudo)scalar boson $\phi$ decaying into $\tau\tau$ pairs, with local(global) significance of $2.6\sigma (2.3\sigma)$ around the mass of 95 GeV. However, attempts to attribute the di-tau excess to a CP-even resonance encounter difficulties, notably due to CMS searches for a scalar resonance in \(t\bar{t}\)-associated production decaying into tau pairs, which do not support such a finding \cite{CMS:2022arx}.

Previously, the Large Electron Positron (LEP) collider collaborations~\cite{LEPWorkingGroupforHiggsbosonsearches}
explored the low-mass domain extensively in the $e^+e^-\to Z\phi$ production mode, with a generic Higgs boson state $\phi$ decaying via the 
$\tau\tau$ and $b\overline{b}$ channels. Interestingly, an excess has been 
reported in 2006
in the $e^+e^- \to Z {\phi}(\to b\overline{b})$ mode for  $m_{b\bar b}$ around 98 GeV~\cite{ALEPH:2006tnd}.
Given the limited mass resolution of the di-jet invariant mass at LEP, this anomaly may well coincide with the aforementioned excesses seen by CMS and/or ATLAS in the $\gamma\gamma$ and $\tau\tau$ final state. Since the excesses appear in very similar mass regions, several studies~\cite{Cao:2016uwt,Heinemeyer:2021msz,Biekotter:2021qbc,Biekotter:2019kde,Cao:2019ofo,Biekotter:2022abc,Iguro:2022dok,Li:2022etb,Cline:2019okt,Biekotter:2021ovi,Crivellin:2017upt,Cacciapaglia:2016tlr,Abdelalim:2020xfk,Biekotter:2022jyr,Biekotter:2023jld,Azevedo:2023zkg,Biekotter:2023oen,Cao:2024axg,Wang:2024bkg,Li:2023kbf,Dev:2023kzu,Borah:2023hqw,Cao:2023gkc,Ellwanger:2023zjc,Aguilar-Saavedra:2023tql,Ashanujjaman:2023etj,Dutta:2023cig,Ellwanger:2024txc,Diaz:2024yfu,Ellwanger:2024vvs,Ayazi:2024fmn,Coloretti:2023wng,Bhattacharya:2023lmu,Ahriche:2023hho,Ahriche:2023wkj} have explored the possibility of simultaneously explaining these anomalies within BSM frameworks featuring a non-standard Higgs state lighter than 125 GeV, while being in agreement with current measurements of the properties of the $\approx 125$ GeV SM-like Higgs state observed at the LHC. In the attempt to explain the excesses in the $\gamma\gamma$ and $b\overline{b}$ channels, it was found in Refs.~\cite{Benbrik:2022azi, Benbrik:2022tlg, Belyaev:2023xnv} that the 2HDM Type-III with a particular Yukawa texture can
successfully accommodate both measurements simultaneously with the lightest CP-even Higgs boson of the model, while being consistent with all relevant theoretical and experimental constraints. 
Further recent studies have shown that actually all three aforementioned signatures can be simultaneously explained in the 2HDM plus a real (N2HDM)~\cite{Biekotter:2022jyr} and complex (S2HDM) ~\cite{Biekotter:2023jld,Biekotter:2023oen} singlet.

In this study, we demonstrate that a superposition of CP-even and CP-odd resonances within the 2HDM Type-III offers a compelling explanation for these excesses at 1$\sigma$ Confidence Level (C.L.) through a $\chi^2$ analysis while, again, satisfying both  theoretical requirements and up-to-date experimental constraints.

The organization of the paper is as follows. Section~\ref{sec:model} reviews the theoretical framework of the 2HDM Type-III, emphasizing its potential in explaining the observed excesses. Section~\ref{sec:excess} provides a detailed account of the excesses, setting the stage for our analysis. In Section~\ref{sec:cons}, we discuss the theoretical and experimental constraints that shape our exploration of the 2HDM Type-III parameter space. Section~\ref{sec:results} details our numerical approach and the outcomes of scanning the 2HDM Type-III parameter space, with the aim of finding plausible explanations for the observed anomalies. We conclude in Section~\ref{sec:con}, underscoring the importance of our findings and their implications for future LHC searches.
\section{2HDM Type-III}\label{sec:model}
The 2HDM serves as one of the most straightforward extensions of the SM. It comprises two complex doublets of Higgs fields, denoted as $\Phi_i$ ($i = 1, 2$), each with a hypercharge of $Y = +1$. The scalar potential, invariant under the SU(2)$_L \otimes$U(1)$_Y$ gauge symmetry, can be expressed as \cite{Branco:2011iw}:
\begin{align}
\mathcal{V} &= m_{11}^2 \Phi_1^\dagger \Phi_1+ m_{22}^2\Phi_2^\dagger\Phi_2 - \left[m_{12}^2
\Phi_1^\dagger \Phi_2 + \rm{H.c.}\right]  ~\nonumber\\&+ \lambda_1(\Phi_1^\dagger\Phi_1)^2 +
\lambda_2(\Phi_2^\dagger\Phi_2)^2 +
\lambda_3(\Phi_1^\dagger\Phi_1)(\Phi_2^\dagger\Phi_2)  ~\nonumber\\ &+
\lambda_4(\Phi_1^\dagger\Phi_2)(\Phi_2^\dagger\Phi_1) +
\frac12\left[\lambda_5(\Phi_1^\dagger\Phi_2)^2 +\rm{H.c.}\right] 
~\nonumber\\&+\left\{\left[\lambda_6(\Phi_1^\dagger\Phi_1)+\lambda_7(\Phi_2^\dagger\Phi_2)\right]
(\Phi_1^\dagger\Phi_2)+\rm{H.c.}\right\} \label{C2HDMpot}
\end{align}
The hermiticity of this potential implies that the parameters $m_{11}^2$, $m_{22}^2$ and $\lambda_{1,2,3,4}$ are real. In contrast, $\lambda_{5,6,7}$ and $m_{12}^2$ can be complex, although they are considered real in the CP-conserving versions of the 2HDM, which we do here as well. Notably, the $\lambda_{6,7}$ terms have a minimal effect in this study and are thus set to zero. This simplification leaves the model with seven independent parameters, reduced to six in our analysis with the assumption of $H$ being  the observed SM-like Higgs boson with a mass of 125 GeV.

The Yukawa sector of the 2HDM involves general scalar-to-fermion couplings, expressed as:
\begin{align}
-{\cal L}_Y &= \bar Q_L Y^u_1 U_R \tilde \Phi_1 + \bar Q_L Y^{u}_2 U_R
\tilde \Phi_2  + \bar Q_L Y^d_1 D_R \Phi_1 
\nonumber \\&+ \bar Q_L Y^{d}_2 D_R \Phi_2 
+  \bar L Y^\ell_1 \ell_R \Phi_1 + \bar L Y^{\ell}_2 \ell_R \Phi_2 + H.c. 
\label{eq:Yu}
\end{align}
Before Electro-Weak Symmetry Breaking (EWSB), the Yukawa matrices $Y^{f}_{1,2}$, which govern the interactions between the Higgs fields and fermions, are arbitrary $3\times 3$ matrices. In this state, fermions do not yet represent physical eigenstates. This allows us the flexibility to choose diagonal forms for the matrices $Y^u_1$, $Y^d_2$ and $Y^\ell_2$. Specifically, we can set $Y^u_1 = \mathrm{diag}(y^u_1, y^u_2, y^u_3)$ and $Y^{d,\ell}_{2} = \mathrm{diag}(y^{d,\ell}_1, y^{d,\ell}_2, y^{d,\ell}_{3})$.

In our study, we focus on the 2HDM Type-III. This variant does not impose a global symmetry on the Yukawa sector nor enforces alignment in flavor space. Instead, we adopt the Cheng-Sher ansatz \cite{Cheng:1987rs, Diaz-Cruz:2004wsi}, which posits a specific flavor symmetry in the Yukawa matrices. Under this assumption, FCNC effects are proportional to the masses of the fermions and dimensionless real parameters \cite{Hernandez-Sanchez:2012vxa} $\chi_{ij}^{f}$ ($\propto \sqrt{m_i m_j}/ v ~\chi_{ij}^f$), where  $i,j=1-3$.
After EWSB, the Yukawa Lagrangian is expressed in terms of the mass eigenstates of the Higgs bosons. It can be represented as follows:
\begin{widetext}
	\begin{align}
	-{\cal L}^{III}_Y  &= \sum_{f=u,d,\ell} \frac{m^f_j }{v} \times\left( (\xi^f_h)_{ij}  \bar f_{Li}  f_{Rj}  h + (\xi^f_H)_{ij} \bar f_{Li}  f_{Rj} H - i (\xi^f_A)_{ij} \bar f_{Li}  f_{Rj} A \right)\nonumber\\  &+ \frac{\sqrt{2}}{v} \sum_{k=1}^3 \bar u_{i} \left[ \left( m^u_i  (\xi^{u*}_A)_{ki}  V_{kj} P_L+ V_{ik}  (\xi^d_A)_{kj}  m^d_j P_R \right) \right] d_{j}  H^+ + \frac{\sqrt{2}}{v}  \bar \nu_i  (\xi^\ell_A)_{ij} m^\ell_j P_R \ell_j H^+ + H.c.\, \label{eq:Yukawa_CH}
	\end{align} 	
\end{widetext}
Here, $V_{kj}$ represents the Cabibbo-Kobayashi-Maskawa (CKM) matrix, while the specific reduced Yukawa couplings are elaborated in Tab. \ref{coupIII}, with expressions defined in relation to the mixing angle $\alpha$, $\tan\beta$ and the independent parameters $\chi_{ij}^f$.
\begin{table*}[t]
	\begin{center}
		\renewcommand{\arraystretch}{0.9} %
			\begin{tabular*}{1.5\columnwidth}{cccccccccc} \noalign{\hrule height 0.9pt}	
			\noalign{\vspace{1.35pt}}  
			\noalign{\hrule height 0.4pt}
			$\phi$  &&& $(\xi^u_{\phi})_{ij}$ &&&  $(\xi^d_{\phi})_{ij}$ &&&  $(\xi^\ell_{\phi})_{ij}$  \\   \noalign{\hrule height 0.9pt}	
			$h$~ 
			&&& ~ $  \frac{c_\alpha}{s_\beta} \delta_{ij} -  \frac{c_{\beta-\alpha}}{\sqrt{2}s_\beta}  \sqrt{\frac{m^u_i}{m^u_j}} \chi^u_{ij}$~
			&&& ~ $ -\frac{s_\alpha}{c_\beta} \delta_{ij} +  \frac{c_{\beta-\alpha}}{\sqrt{2}c_\beta} \sqrt{\frac{m^d_i}{m^d_j}}\chi^d_{ij}$~
			&&& ~ $ -\frac{s_\alpha}{c_\beta} \delta_{ij} + \frac{c_{\beta-\alpha}}{\sqrt{2}c_\beta} \sqrt{\frac{m^\ell_i}{m^\ell_j}}  \chi^\ell_{ij}$ ~ \\
			$H$~
			&&& $ \frac{s_\alpha}{s_\beta} \delta_{ij} + \frac{s_{\beta-\alpha}}{\sqrt{2}s_\beta} \sqrt{\frac{m^u_i}{m^u_j}} \chi^u_{ij} $
			&&& $ \frac{c_\alpha}{c_\beta} \delta_{ij} - \frac{s_{\beta-\alpha}}{\sqrt{2}c_\beta} \sqrt{\frac{m^d_i}{m^d_j}}\chi^d_{ij} $ 
			&&& $ \frac{c_\alpha}{c_\beta} \delta_{ij} -  \frac{s_{\beta-\alpha}}{\sqrt{2}c_\beta} \sqrt{\frac{m^\ell_i}{m^\ell_j}}  \chi^\ell_{ij}$ \\
			$A$~  
			&&& $ \frac{1}{t_\beta} \delta_{ij}- \frac{1}{\sqrt{2}s_\beta} \sqrt{\frac{m^u_i}{m^u_j}} \chi^u_{ij} $  
			&&& $ t_\beta \delta_{ij} - \frac{1}{\sqrt{2}c_\beta} \sqrt{\frac{m^d_i}{m^d_j}}\chi^d_{ij}$  
			&&& $t_\beta \delta_{ij} -  \frac{1}{\sqrt{2}c_\beta} \sqrt{\frac{m^\ell_i}{m^\ell_j}}  \chi^\ell_{ij}$ \\\noalign{\hrule height 0.4pt} 	
			\noalign{\vspace{1.35pt}}  
			\noalign{\hrule height 0.9pt}	
		\end{tabular*}
	\end{center}
	\caption {Yukawa couplings of the neutral Higgs bosons $h$, $H$, and $A$ to the quarks and leptons in the 2HDM Type-III. Off-diagonal elements $\chi^{f}_{ij}$ (for $i \neq j$) are set to zero.} 
	\label{coupIII}
\end{table*}
\section{The Excesses in the $\gamma\gamma$, $\tau\tau$ and $b\bar b$ Channels}\label{sec:excess}

In this section, we investigate whether the 2HDM Type-III can describe consistently the excesses observed by both LEP and the LHC  in the 94--100 GeV  mass windows in the $b\bar b$ as well as $\gamma\gamma$ and $\tau\tau$  channels, respectively. Starting with the LHC excesses, the parametrization used to access  possible BSM signals invokes the so-called `signal strength' (defined in terms of ratios of production cross sections $\sigma$ and decay Branching Ratios ${\cal BR}$s), which,  for these excesses, are  as follows:

\begin{eqnarray}
\mu_{\mathrm{\tau\tau}}&=&\frac{\sigma_{\rm 2HDM}(gg\to \phi )}{\sigma_{\rm SM}(gg\to h_{\rm SM})}\times \frac{{\cal BR}_{\rm 2HDM}(\phi \to \tau\tau)}{{\cal BR}_{\rm SM}(h_{\rm SM}\to \tau\tau)},\nonumber\\
\mu_{\mathrm{\gamma\gamma}}&=&\frac{\sigma_{\rm 2HDM}(gg\to \phi )}{\sigma_{\rm SM}(gg\to h_{\rm SM})}\times \frac{{\cal BR}_{\rm 2HDM}(\phi \to \gamma\gamma)}{{\cal BR}_{\rm SM}(h_{\rm SM}\to \gamma\gamma)}.\label{mu_cms}
\end{eqnarray} 	

The experimental measurements for these two signal strengths are expressed as \cite{Biekotter:2023oen,Biekotter:2022jyr,Biekotter:2023jld}:
\begin{eqnarray}
\mu_{\gamma\gamma}^{\mathrm{exp}}&=&\mu_{\gamma\gamma}^{\mathrm{ATLAS+CMS}} = 0.24^{+0.09}_{-0.08},\\
\mu_{\tau\tau}^{\mathrm{exp}} &=& 1.2 \pm 0.5,
\end{eqnarray}
where $h_{\rm SM}$ refers to a hypothetical SM Higgs boson assumed at a mass of $\sim 95$ GeV.

In our analysis, we have combined the di-photon measurements from the ATLAS and CMS experiments, denoted as $\mu_{\gamma\gamma}^{\mathrm{ATLAS}}$ and $\mu_{\gamma\gamma}^{\mathrm{CMS}}$, respectively. The ATLAS measurement yields a central value of $0.18{\pm0.1}$ \cite{Biekotter:2023oen} while the CMS measurement yields a central value of $0.33^{+0.19}_{-0.12}$\cite{Biekotter:2023jld}. By doing so, we aimed to leverage the strengths of both experiments and improve the precision of our analysis. The combined measurement, denoted as $\mu_{\gamma\gamma}^{\mathrm{ATLAS+CMS}}$, is determined by taking the average of the central values without assuming any correlation between them. To evaluate  the combined uncertainty we sum ATLAS and CMS uncertainties in quadrature.

The signal strength for the $b\bar{b}$ channel from LEP data is defined as:
\begin{eqnarray}\label{mu_lep}
\mu_{{b\bar{b}}} &=& \frac{\sigma_{\mathrm{2HDM}}(e^+e^- \to Zh)}{\sigma_{\mathrm{SM}}(e^+e^- \to Zh_{\mathrm{SM}})} \times \frac{\mathcal{BR}_{\mathrm{2HDM}}(h \to b\bar{b})}{\mathcal{BR}_{\mathrm{SM}}(h_{\mathrm{SM}} \to b\bar{b})}. 
\end{eqnarray}
Here, the expected experimental value of \(\mu_{b\bar{b}}\) is reported as \cite{LEPWorkingGroupforHiggsbosonsearches}:
\begin{eqnarray}
\mu_{b\bar{b}}^{\mathrm{exp}} = 0.117 \pm 0.057.
\end{eqnarray}

To determine whether a simultaneous fit to the observed excesses is possible, a $\chi^2$ analysis is performed using the measured central values $\mu^{\mathrm{exp}}$ and the 1$\sigma$ uncertainties $\Delta\mu^{\mathrm{exp}}$ of the signal rates related to the three excesses as defined in Eqs.~(\ref{mu_cms}) and (\ref{mu_lep}). The contribution to the $\chi^2$ value for each  channel  is calculated using the formula:
\begin{eqnarray}
\chi^2_{\gamma\gamma,\tau\tau(+b\bar b)}=\frac{\left(\mu_{\gamma\gamma,\tau\tau(+b\bar b)}-\mu_{\gamma\gamma,\tau\tau(+b\bar b)}^\mathrm{ exp}\right)^2}{\left(\Delta\mu^\mathrm{exp}_{\gamma\gamma,\tau\tau(+b\bar b)}\right)^2}.
\end{eqnarray}
{So, the resulting $\chi^2$, 
	which we will use to judge whether the points from the model describe the 
	excesses, reads as:
}
\begin{equation}
\chi^2_{\gamma\gamma+\tau\tau(+b\bar b)}=
\chi^2_{\gamma\gamma}+\chi^2_{\tau\tau}~(+\chi^2_{b\bar b}),
\label{eq:chi95}
\end{equation}
where the inclusion of the $b\bar b$ data depends on the solution that we will attempt. Specifically, having tested the $h$-only explanation in Refs.~\cite{Benbrik:2022azi,Benbrik:2022tlg,Belyaev:2023xnv}, here, we are concerned with the $A$-only one (which will then necessarily not capture the LEP data as there is no $AZZ$ coupling) as well as with the superposition of the two (which can potentially capture both LHC and LEP anomalies). However,  
before using the above measure to test the viability of  2HDM Type-III  against the anomalous LHC data, we describe the aforementioned theoretical and experimental constraints adopted here.

\section{Theoretical and experimental constraints}\label{sec:cons}
In our work, we employ a diverse set of theoretical and experimental constraints that must be met to establish a viable model.
\begin{itemize}
	\item \textbf{Unitarity} The scattering processes involving (pseudo)scalar-(pseudo)scalar, gauge-gauge and/or (pseudo)scalar-gauge initial and/or final states must satisfy unitarity constraints. The eigenvalues $e_i$ of the tree-level 2-to-2 body scattering matrix should meet the following criteria: $|e_i|<8\pi$ ~\cite{uni1,uni2}.
	
	\item \textbf{Perturbativity} Adherence to perturbativity constraints imposes an upper limit on the quartic couplings of the Higgs potential: $|\lambda_i|<8\pi$ ~\cite{Branco:2011iw}.
	
	\item \textbf{Vacuum Stability} The scalar potential must be positive and bounded from below in any direction of the fields $\Phi_i$ to ensure vacuum stability. This requires that $\lambda_1>0$, $\lambda_2>0$, $\lambda_3>-\sqrt{\lambda_1\lambda_2}$, and $\lambda_3+\lambda_4-|\lambda_5|>-\sqrt{\lambda_1\lambda_2}$ ~\cite{Barroso:2013awa,sta}.
	\begin{align}
	&\lambda_1 > 0, \quad \lambda_2 > 0, \quad \lambda_3 > -\sqrt{\lambda_1\lambda_2}, \nonumber\\
	&\lambda_3+\lambda_4-|\lambda_5| > -\sqrt{\lambda_1\lambda_2}.
	\end{align}
	\item \textbf{SM-like Higgs Boson Discovery} The compatibility of the SM-like scalar with the observed Higgs boson is tested. The relevant quantities calculated with \texttt{HiggsSignals-3} \cite{Bechtle:2020pkv,Bechtle:2020uwn} via \texttt{HiggsTools} \cite{Bahl:2022igd}  must satisfy the measurements at 95\% confidence level (C.L.).
	\item \textbf{BSM Higgs Boson Exclusions} Exclusion limits at 95\% C.L. from direct searches for Higgs bosons at LEP, Tevatron, and LHC are taken into account using \texttt{HiggsBounds-6} \cite{Bechtle:2008jh,Bechtle:2011sb,Bechtle:2013wla,Bechtle:2015pma} via \texttt{HiggsTools}.
	
	\item
	{\bf $B$-physics observables} The constraints from $B$-physics observables are implemented using the code \texttt{SuperIso\_v4.1} \cite{superIso} as described in Ref. \cite{Benbrik:2022azi}. The relevant experimental measurements used are as follows:
	
	{	\begin{enumerate}
			\item ${\cal BR}(\overline{B}\to X_s\gamma)|_{E_\gamma<1.6\mathrm{~GeV}}$ $\left(3.32\pm0.15\right)\times 10^{-4}$ \cite{HFLAV:2016hnz},
			\item${\cal BR}(B^+\to \tau^+\nu_\tau)$ $\left(1.06\pm0.19\right)\times 10^{-4}$ \cite{HFLAV:2016hnz},
			\item${\cal BR}(D_s\to \tau\nu_\tau)$ $\left(5.51\pm0.18\right)\times 10^{-2}$ \cite{HFLAV:2016hnz},
			\item${\cal BR}(B_s\to \mu^+\mu^-)$ (LHCb) $\left(3.09^{+0.46}_{-0.43}\right)\times 10^{-9}$ \cite{LHCb:2021awg,LHCb:2021vsc},
			\item	${\cal BR}(B_s\to \mu^+\mu^-)$ (CMS) $\left(3.83^{+0.38}_{-0.36}\right)\times 10^{-9}$ \cite{CMS:2022mgd},
			\item	${\cal BR}(B^0\to \mu^+\mu^-)$ (LHCb) $\left(1.2^{+0.8}_{-0.7}\right)\times 10^{-10}$ \cite{LHCb:2021awg,LHCb:2021vsc},
			\item	${\cal BR}(B^0\to \mu^+\mu^-)$ (CMS) $\left(0.37^{+0.75}_{-0.67}\right)\times 10^{-10}$ \cite{CMS:2022mgd}.
			%\item	${\cal BR}(K\to \mu\nu_\mu)/{\cal BR}(\pi\to \mu\nu_\mu)$ $0.6357 \pm 0.0011$ \cite{LHCb:2017rmj}.
	\end{enumerate} }
\end{itemize}

%%%%%%%%%%%%%%%%%%%%%%%%%%%%%%%%%%%%%%%%%%%%%%%%%%%%%%%%%%%     		
\section{Explanation of the Excesses}\label{sec:results}
In this section, we present  our numerical analysis of the 
2HDM Type-III parameter space. For the 2HDM Type-III spectrum generation, we have employed \texttt{2HDMC}  
\cite{2HDMC}, which considers the theoretical constraints discussed in the previous section, along with the Electro-Weak Precision Observables (EWPOs). Subsequently, we validate our results by comparing them to Higgs data, utilizing \texttt{HiggsTools} \cite{Bahl:2022igd}, which includes the most recent versions of both \texttt{HiggsBounds} and \texttt{HiggsSignals}.
In accordance with the above discussions,
we consider the scenario where the heavier CP-even Higgs boson $H$ is the SM-like Higgs
particle $H_{\rm SM}$ discovered at the LHC with $m_{H_{\rm SM}}\approx$  125 GeV. In this scenario the CP-odd Higgs, $A$, is the source of the observed LHC excess
in $\gamma\gamma$ and  $\tau\tau$  channels around 95 GeV, which we previously labelled as $h_{\rm SM}$.
To explore this scenario, we conducted a systematic random scan across the parameter ranges specified in Tab.~\ref{tab:par-scan}.

\begin{table}[H]
	\centering
	%		{\renewcommand{\arraystretch}{1.55} %donne la distance entre les lignes%
	{\setlength{\tabcolsep}{1.0cm}
		\begin{tabular*}{\columnwidth}{ccc}
		\noalign{\hrule height 0.9pt}
		\noalign{\vspace{1.35pt}}  
		\noalign{\hrule height 0.4pt}
			Parameters  && Scanned ranges \\
		\noalign{\hrule height 0.9pt}
	
			$m_h$   && [$94$, $97$] \\
			$m_H$  && $125.09$ \\
			$m_A$  && [$94$, $97$] \\
			$m_{H^\pm}$  && [$160$, $300$] \\
			$\tan\beta$ && [$1$, $10$] \\
			$s_{\beta-\alpha}$  && [-$0.5$, $0$] \\	
			$\chi_{ij}^{f,\ell} $  && [$-3$, $3$] \\
		\noalign{\hrule height 0.4pt}
			\noalign{\vspace{1.35pt}}  
				\noalign{\hrule height 0.9pt}
	\end{tabular*}}
	\caption{Scan ranges of the 2HDM Type-III input parameters. Masses are given in GeV.}
	\label{tab:par-scan}
\end{table}
\noindent
%%%%%%%%%%%%%%%%%%%%%%%%%%%%%%%%%% comment on Fig 1 
\subsection{The $A$ Solution}
Here, we  investigate parameter spaces that satisfy the condition $\chi^2_{125}\le 189.4$, corresponding to a 95\% C.L. for 159 degrees of freedom, where $\chi^2_{125}$ corresponds to the $\chi^2$ evaluated by \texttt{HiggsSignals} for the 125 GeV Higgs signal strength measurements. Subsequently, we examine 2-dimensional (2D) planes of the signal strength parameters: $(\mu_{\gamma\gamma}-\mu_{\tau\tau})$.
\begin{figure}[htp!]	
	\centering
	\mbox{\includegraphics[height=11cm,width=0.975\columnwidth]{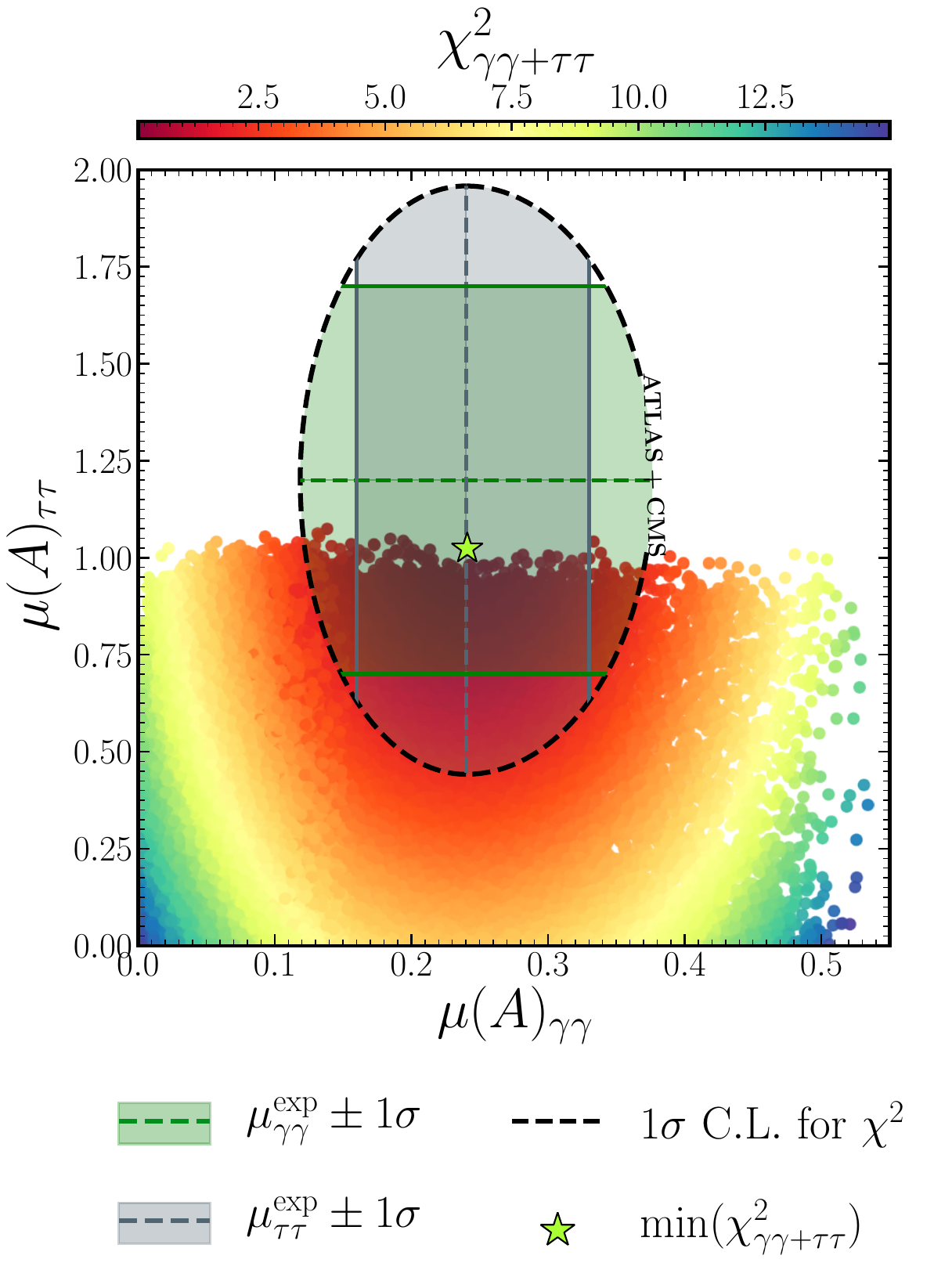}}
	\caption{Di-photon signal rate $\mu(A)_{\gamma\gamma}$ versus di-tau signal rate $\mu(A)_{\tau\tau}$  for a 95 GeV pseudo-scalar within the 2HDM Type-III. The bands represent the experimentally observed signal rates along with their 1$\sigma$ uncertainty intervals. The minimum value of $\chi^2_{\gamma\gamma+\tau\tau}$ is marked by a green star, with a value of 0.12. The black dashed line delineates the 1$\sigma$ C.L. region with respect to $\chi^2_{\gamma\gamma+\tau\tau}$.}\label{fig1}
\end{figure}

In Fig.~\ref{fig1}, we present the results for $\chi^2_{\gamma\gamma+\tau\tau}$ in the form of a color map projected onto the ($\mu_{\tau\tau}-\mu_{\gamma\gamma}$) plane, representing the signal strength parameters. The dashed ellipse delineates the regions consistent with the excess observed at the 1$\sigma$ C.L., as described by the equation $\chi^2_{\gamma\gamma}+\chi^2_{\tau\tau}=2.30$. The value of $\chi^2_{\gamma\gamma+\tau\tau}$ is represented by the vertical color map. The gray(green) dashed line represents the central value for $\mu_{\gamma\gamma}$($\mu_{\tau\tau}$)
the gray(green) band showing the 1$\sigma$ range. The green star indicates the position of $\chi^2_{\gamma\gamma+\tau\tau, \mathrm{min}}$, which is the minimum value of $\chi^2_{\gamma\gamma+\tau\tau}$, noted at 0.12. Furthermore, numerous points surrounding $\chi^2_{\gamma\gamma+\tau\tau, \mathrm{min}}$ are depicted in dark orange, demonstrating the capability of the 2HDM Type-III model with a CP-odd resonance to perfectly explain the observed excess across both channels simultaneously, as well as individually, at the 1$\sigma$  level.

\begin{figure}[htp!]	
	\centering
	\mbox{\includegraphics[height=7cm,width=0.925\columnwidth]{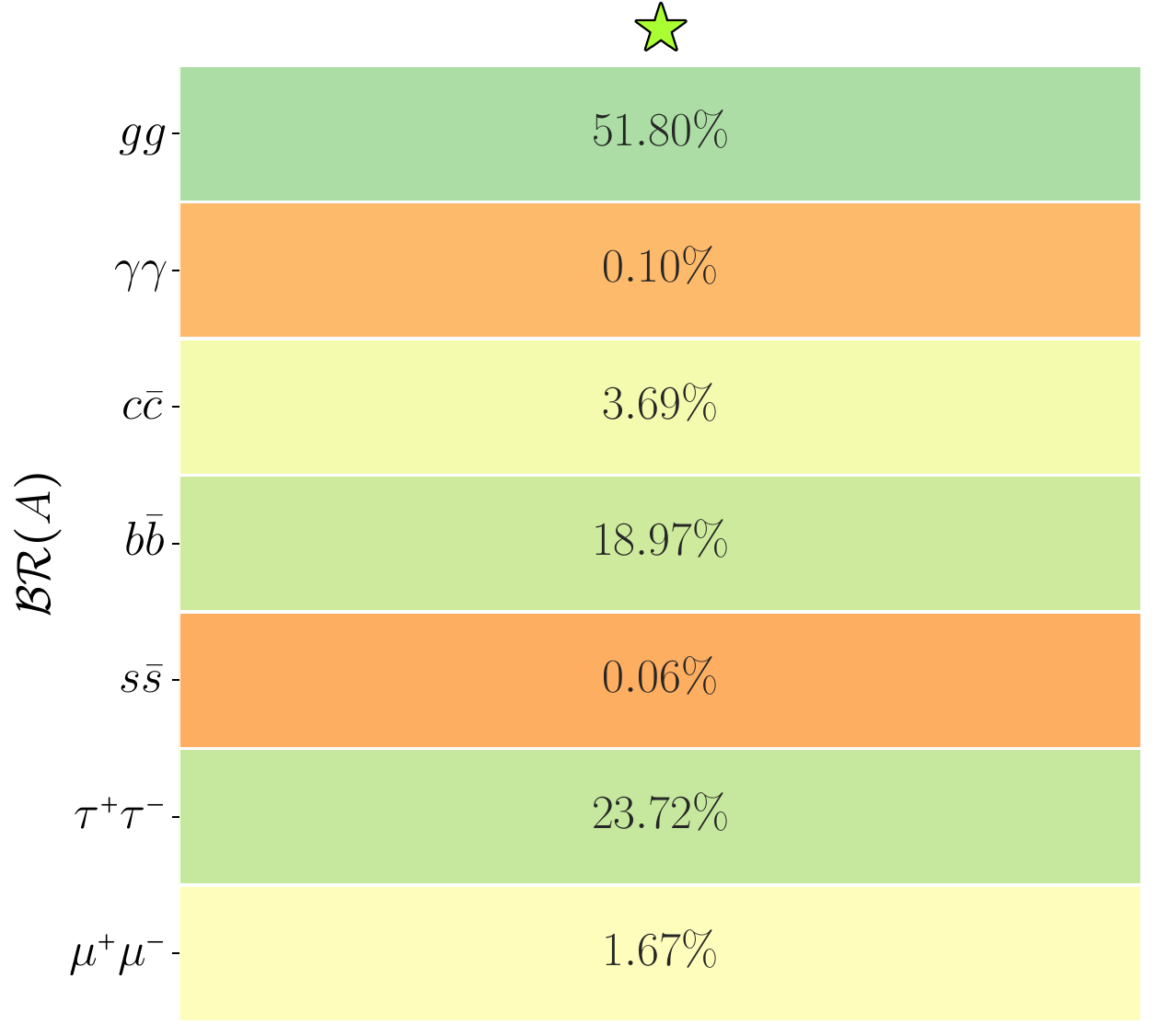}}
	\caption{$\mathcal{BR}s$ of the CP-odd Higgs boson $A$ at our best fit point.}\label{fig2}
\end{figure}

Fig.~\ref{fig2} depicts the values of the branching ratios, $\mathcal{BR}(A)$, for our best-fit point through various possible decay channels.

\begin{figure}[H]	
	\centering
	\mbox{\includegraphics[height=10.5cm,width=\columnwidth]{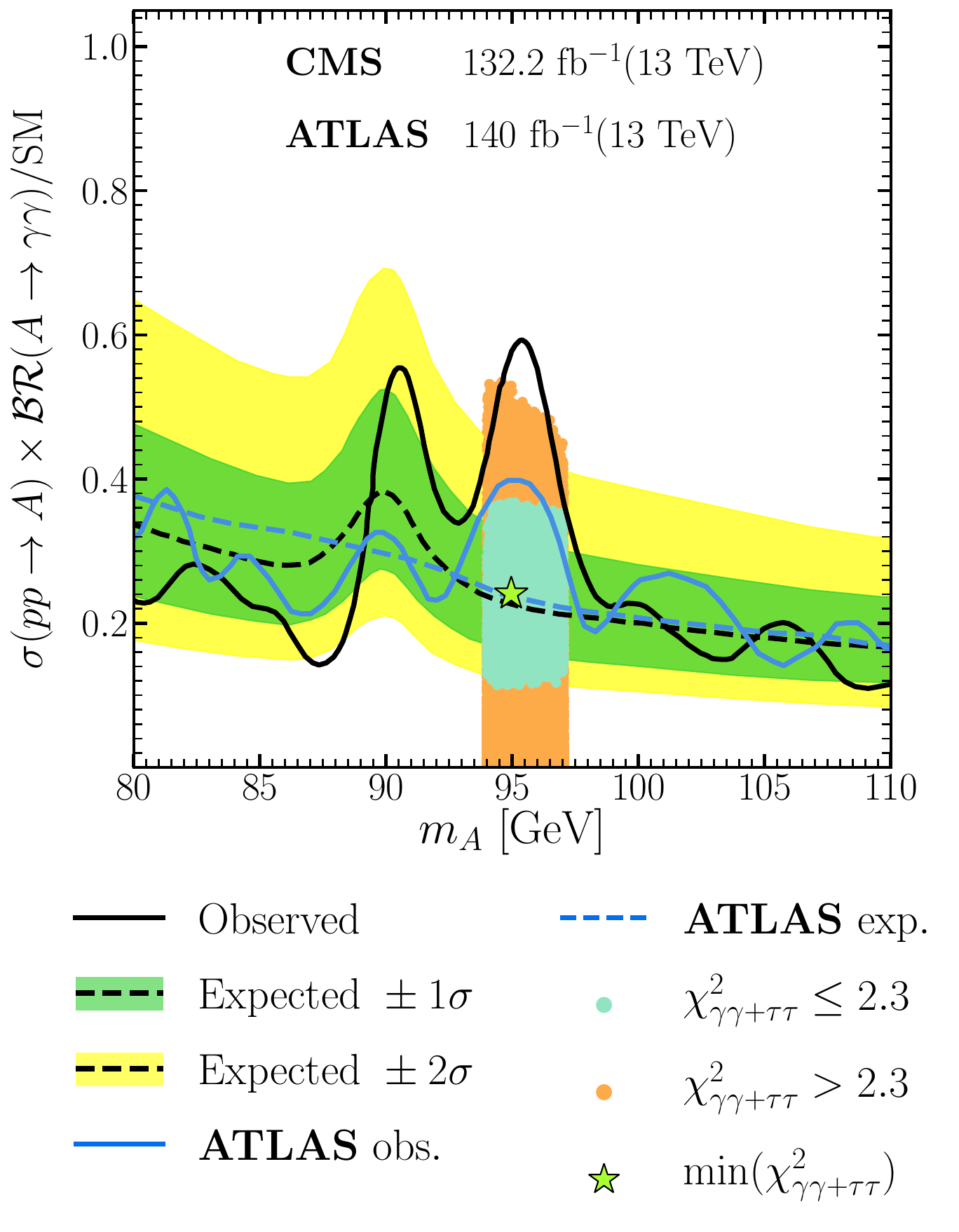}}
	\caption{Allowed points, following the discussed theoretical and experimental constraints,
		superimposed onto the results of the CMS 13 TeV low-mass  $\gamma\gamma$  \cite{CMS:2023yay}  analysis. {(Notably, the plot further includes the depiction of the ATLAS expected and observed limits from \cite{Arcangeletti}, showcased in blue.)}
		The light green colour represents the parameter points that fit the excesses within a
		three-dimensional C.L. of 1$\sigma$,
		whereas the points that fit the excesses at
		2$\sigma$ or more are shown in orange. }\label{fig3}
\end{figure}

\begin{figure}[htp!]	
	\centering
	\mbox{\includegraphics[height=10.5cm,width=\columnwidth]{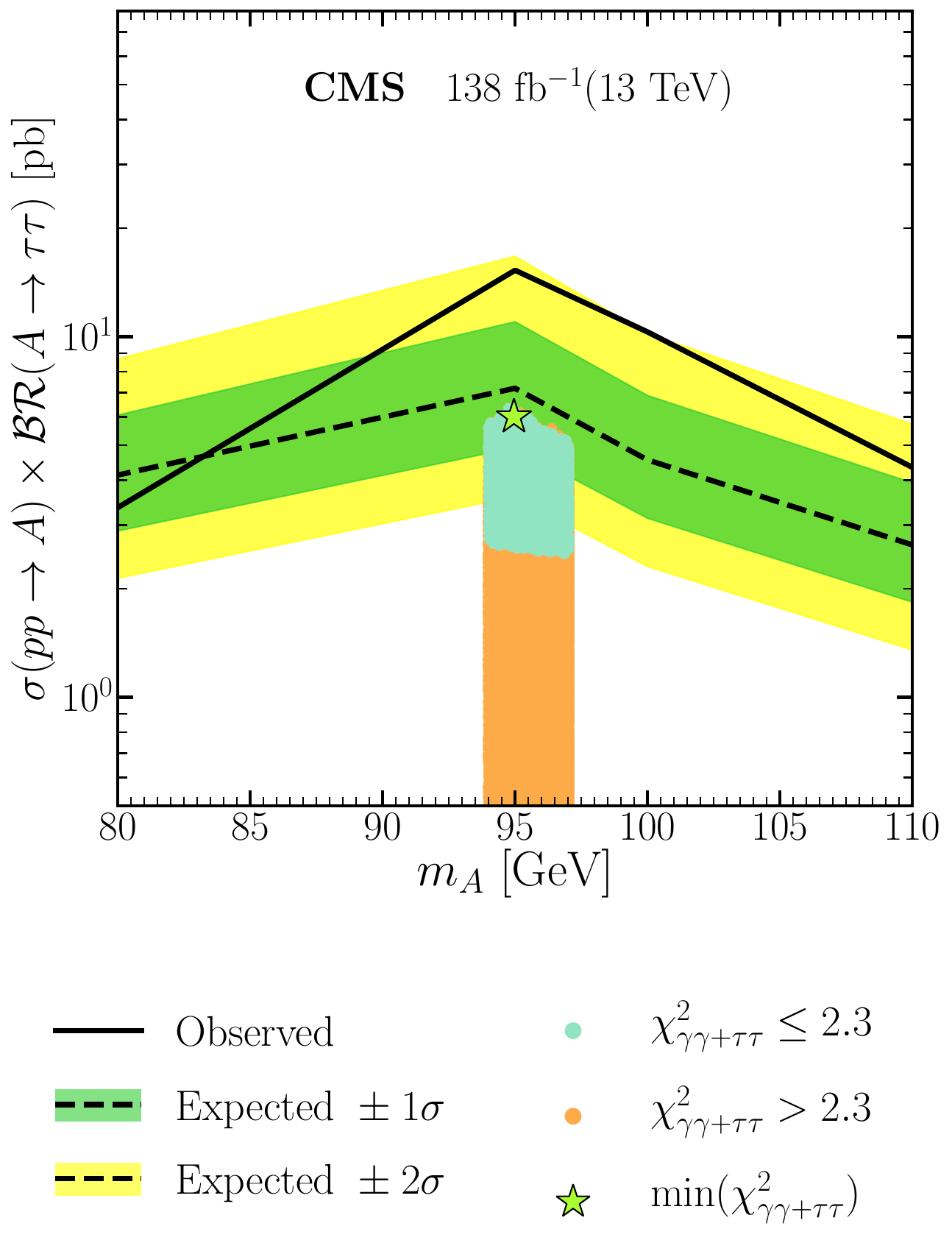}}
	\caption{The same points as in Fig.~\ref{fig3},
		superimposed onto the results of the CMS 13 TeV low-mass  $\tau\tau$~\cite{CMS:2022rbd}  analysis. }\label{fig4}
\end{figure}

In Figs.~\ref{fig3} and \ref{fig4}, we directly compare our allowed parameter points with the experimental data by superimposing them onto the CMS 13 TeV low-mass $\gamma\gamma$ \cite{CMS:2023yay} and $\tau\tau$ \cite{CMS:2022rbd} analysis data, respectively. The light green colour represents the parameter points that fit the excesses within a two-dimensional confidence level (C.L.) of 1$\sigma$, while the points fitting the excesses at 2$\sigma$ or more are shown in orange. It can be clearly observed from the plots that our parameter points are well-suited to satisfy the LHC excesses.

\subsection{The \(h\) + \(A\) Solution}
In this section, we conduct a combined analysis of the CP-even ($h$)\footnote{It should be noted that the CMS \cite{CMS:2022arx} limit for the production of a Higgs boson in association with either a top-quark pair or a $Z$ boson, subsequently decaying into a tau pair, is taken into account in our analysis for the scalar $h$ resonance.} and CP-odd ($A$) resonances. We now explore also the $b\bar{b}$ excess, potentially attributable to the $h$ resonance, by incorporating the $\chi^2_{b\bar{b}}$ into our total $\chi^2$ analysis.

Then, we calculate the combined contributions to the signal strengths from both resonances for the \(\gamma\gamma\) and \(\tau\tau\) channels, as follows:
\begin{eqnarray}
\mu_{\gamma\gamma}(h+A) &=& \mu_{\gamma\gamma}(h) + \mu_{\gamma\gamma}(A), \nonumber \\
\mu_{\tau\tau}(h+A) &=& \mu_{\tau\tau}(h) + \mu_{\tau\tau}(A),
\end{eqnarray}

since there is no interference between the $h$ and $A$ states, given that in our 2HDM Type-III we have assumed CP conservation.

\begin{figure}[H]	
	\centering
	\mbox{\includegraphics[width=\columnwidth]{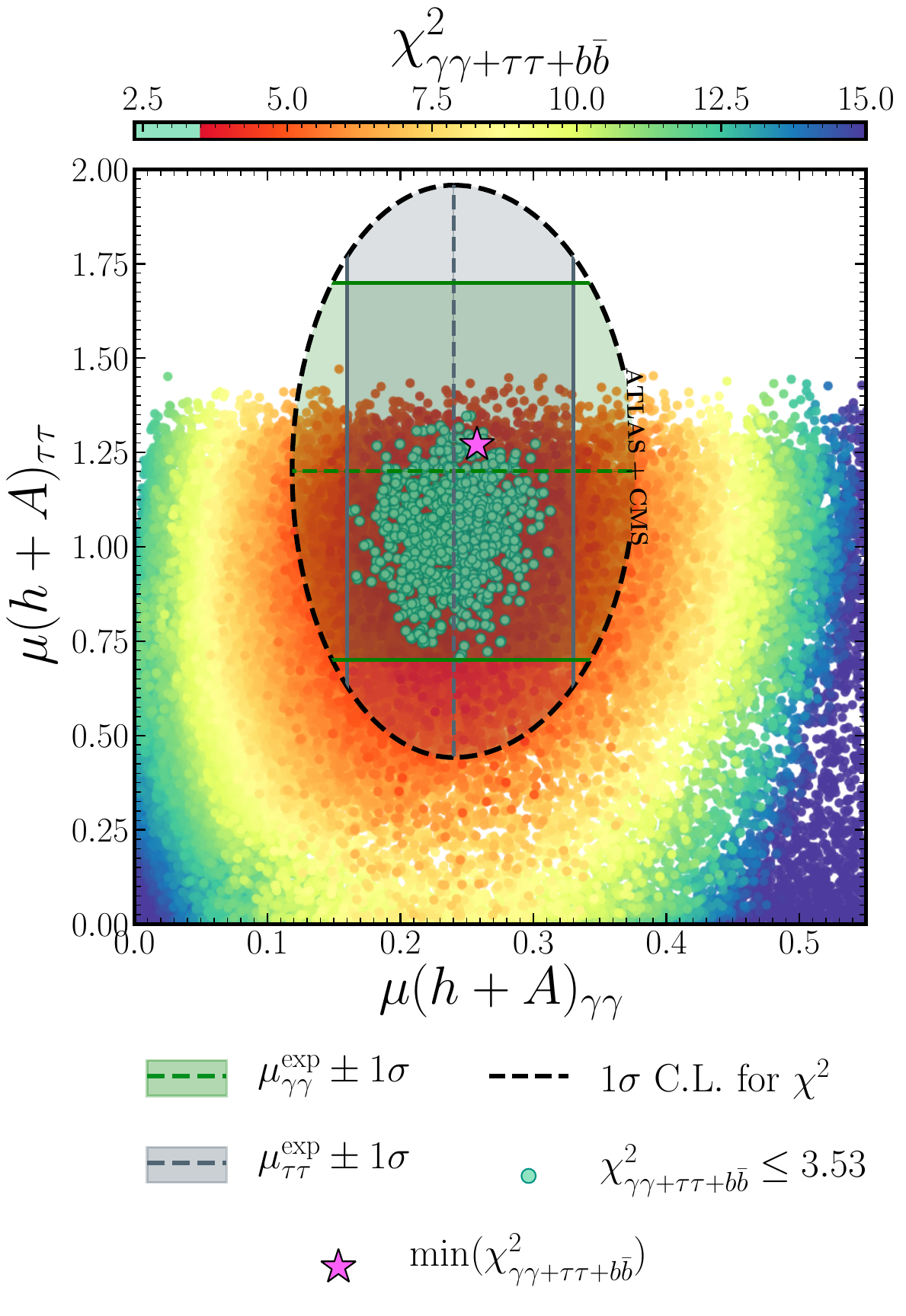}}
	\caption{Di-photon signal rate $\mu(h+A)_{\gamma\gamma}$ versus di-tau signal rate $\mu(h+A)_{\tau\tau}$  for a 95 GeV pseudo-scalar within the 2HDM Type-III. The bands represent the experimentally observed signal rates along with their 1$\sigma$ uncertainty intervals. The minimum value of $\chi^2_{\gamma\gamma+\tau\tau+b\bar b}$ is marked by a magenta star, with a value of 2.35.}\label{fig5}
\end{figure}

Integrating data from both resonances, $h$ and $A$, we demonstrate the 2HDM Type-III ability  to account for observed excesses through their superposition, achieving a 1$\sigma$ C.L. This is clearly illustrated in Fig.~\ref{fig5}, which shows the combined $\chi^2_{\gamma\gamma+\tau\tau+b\bar{b}}$ in the signal strengths $(\mu_{\gamma\gamma}, \mu_{\tau\tau})$ plane. Points that explain the three excesses at 1$\sigma$ ($\chi^2 \leq 3.53$) are marked in green. Notably, the minimum value of $\chi^2_{\gamma\gamma+\tau\tau+b\bar{b}}$ is 2.35, which is highlighted by a magenta star.

\begin{figure}[H]	
	\centering
	\mbox{\includegraphics[width=\columnwidth]{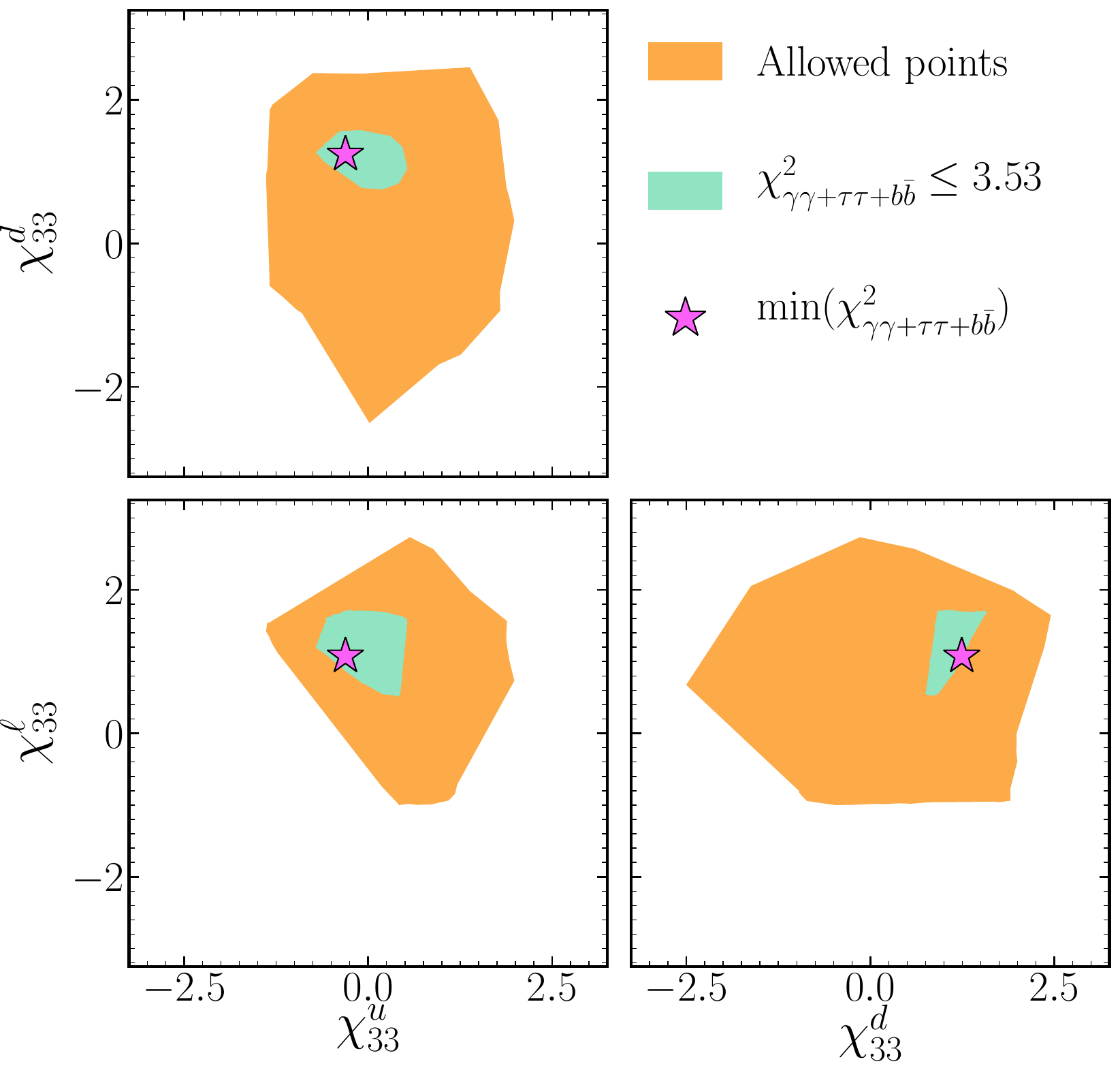}}
	\caption{Allowed points, following the discussed theoretical and experimental constraints (orange), while the area indicating $\chi^2_{\gamma\gamma+\tau\tau+b\bar b}\le 3.53$ (1$\sigma$ C.L) is illustrated in teal. The magenta star marks the minimum of  $\chi^2_{\gamma\gamma+\tau\tau+b\bar b}$ . }\label{fig6}
\end{figure}

Additional insights are provided in Fig.~\ref{fig6}, displaying allowed parameter points in the $(\chi^{u}_{33}-\chi^{\ell}_{33})$, $(\chi^{u}_{33}-\chi^{d}_{33})$, and $(\chi^{d}_{33}-\chi^{\ell}_{33})$ planes, highlighted in orange. Areas meeting the criterion $\chi^2_{\gamma\gamma+\tau\tau+b\bar{b}} \leq 3.53$ (1$\sigma$ C.L.) are indicated in green. The analysis confirms that to accommodate the three excesses at the 1$\sigma$ level, the necessary parameter intervals are: $\chi^{u}_{33} \in [-0.71, 0.53]$, $\chi^{d}_{33} \in [0.75, 1.57]$, and $\chi^{\ell}_{33} \in [0.52, 1.70]$.

\begin{figure}[H]	
	\centering
	\mbox{\includegraphics[width=\columnwidth]{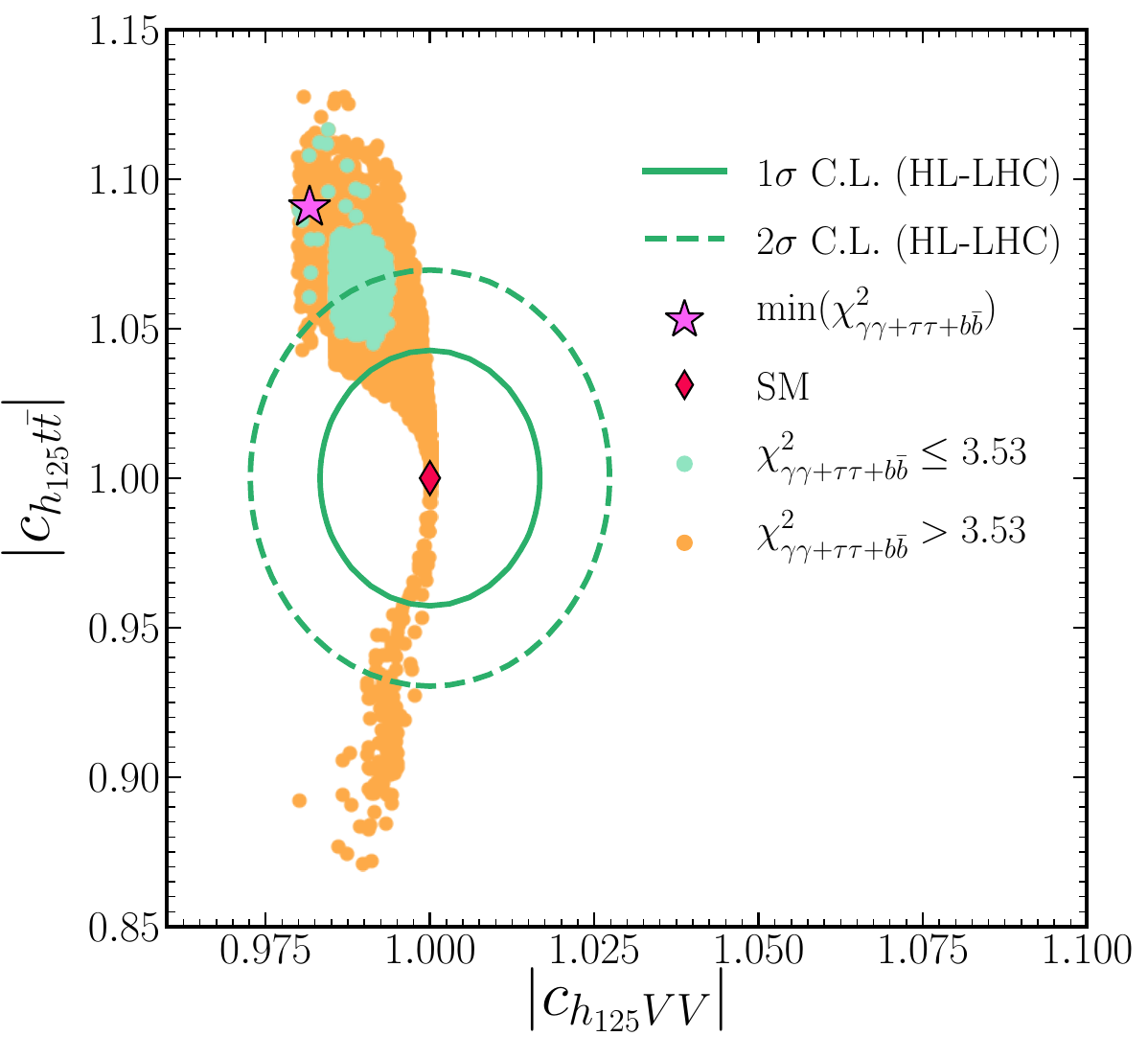}}
	\caption{Correlation between the normalized couplings $|c_{h_{125}VV}|$ and $|c_{h_{125}t\bar t}|$, with colours corresponding to those in Fig.~\ref{fig6}, are illustrated.  The green solid (dashed)  ellipses indicate the projected uncertainties at the HL-LHC \cite{Cepeda:2019klc} at $1\sigma$($2\sigma$). }\label{fig7}
\end{figure}
Fig.~\ref{fig7} illustrates the correlation between the normalized couplings of the $\approx 125$ GeV Higgs, using the same color scheme as detailed in Figure~\ref{fig5}. The plot features green solid and dashed lines, representing the projected experimental precision for these couplings at the High Luminosity LHC (HL-LHC) at the 1$\sigma$ and 2$\sigma$ levels, respectively, based on an expected integrated luminosity of 3000 fb$^{-1}$. The center of these projections, corresponding to the SM values, is marked by a red diamond. Our analysis reveals that explaining the observed excesses in the $b\overline{b}$, $\tau\tau$, and $\gamma\gamma$ channels requires an enhancement of the $h_{125}t\overline{t}$ couplings, which deviate by approximately 12\% from the SM predictions for $t\bar{t}$, as discussed in \cite{Belyaev:2023xnv}. Additionally, it is evident that each point that successfully accounts for the three excesses consistently lies outside the 1$\sigma$ ellipse, though some points may fit within the 2$\sigma$ level. Given these deviations, the expected precision of HL-LHC experiments will enable a clear differentiation between the SM-like properties of $h_{125}$ and those of the $H$ boson from the 2HDM Type-III model within the parameter ranges consistent with these observed excesses.

In summary, this section provides a comprehensive overview of our best fit points, as presented in Tab. \ref{tab:BF}. The first point corresponds to the best fit for the CP-odd state, explaining the LHC excesses in the $\gamma\gamma$ and $\tau\tau$ channels. The second point, indicated by a magenta star, represents the best fit point for the superposition solution of both CP-even and CP-odd resonances, addressing the LHC excesses along with the LEP excess in the $b\bar{b}$ channel.

\begin{table*}[t!] 
		{\footnotesize	
				\setlength{\tabcolsep}{0.2cm}
				\renewcommand*{\arraystretch}{1.9}
				\begin{tabular*}{\textwidth}{ @{\extracolsep{\fill}} lccccccccccccccc}\noalign{\hrule height 0.9pt}
					\noalign{\vspace{1.35pt}}  
					\noalign{\hrule height 0.4pt}
					\textbf{Parameters} &
					
					$m_{h}$ & $m_{H}$ & $m_{A}$ & $m_{H^\pm}$ & $t_{\beta}$ & $s_{\beta-\alpha}$ & $\chi^u_{11}$ &  $\chi^u_{22}$ & $\chi^u_{33}$ & $\chi^d_{11}$ & $\chi^d_{22}$ & $\chi^d_{33}$ & $\chi^\ell_{11}$ & $\chi^\ell_{22}$ & $\chi^\ell_{33}$\\
					\noalign{\hrule height 0.9pt}
				{{\color{lawngreen}\FiveStar\hspace{-1em}\color{black}\textbf{\FiveStarOpen}}}
					&94.62 & 125.09 & 94.96 & 162.95 &   1.82 &   -0.16 &     1.55 &     0.33 &    -0.04 &    -0.10 &     1.56 &     1.14 &     0.67 &    -0.44 &     1.62 \\
					{\color{magenta}\FiveStar\hspace{-1em}\color{black}\textbf{\FiveStarOpen}} & 95.61 & 125.09 & 94.36 & 162.92 &   2.93 &   -0.19 &     0.56 &     0.36 &    -0.31 &    -0.01 &    -0.16 &     1.24 &     1.27 &     0.44 &     1.08\\\noalign{\hrule height 0.4pt}
					\noalign{\vspace{1.25pt}}  \noalign{\hrule height 0.9pt}
				\end{tabular*}
				{	\setlength{\tabcolsep}{0.27cm}
					\begin{tabular*}{\textwidth}{@{\extracolsep{\fill}}lccccccccccccccc}
						\textbf{Signal strengths} &&&
						$\mu_{\gamma\gamma}(h)$ && $\mu_{\gamma\gamma}(A)$ &&$\mu_{\gamma\gamma}(h+A)$ && $\mu_{\tau\tau}(h)$ && $\mu_{\tau\tau}(A)$&&$\mu_{\tau\tau}(h+A)$&&
						$\mu_{b\bar b}(h)$\\
						\noalign{\hrule height 0.9pt}
						{\color{lawngreen}\FiveStar\hspace{-1em}\color{black}\textbf{\FiveStarOpen}}
						&&& 0.24 &&      0.16 &&       0.40 &&      0.22 &&     1.02 &&       1.25 &&    0.02  \\
						\color{magenta}\FiveStar\hspace{-1em}\color{black}\textbf{\FiveStarOpen} &&& 0.18 &&      0.08 &&       0.26 &&      0.38 &&     0.89 &&       1.27 &&    0.03 \\\noalign{\hrule height 0.4pt}
						\noalign{\vspace{1.25pt}}  \noalign{\hrule height 0.9pt}
						
				\end{tabular*}}
				{	\setlength{\tabcolsep}{0.54cm}\begin{tabular*}{\textwidth}{@{\extracolsep{0.5pt}}lccccccccccc}
						\textbf{Effective couplings} &
						$c_{h_{95} t \bar t}$ & $c_{h_{95} b \bar b}$
						& $c_{h_{95} VV} $ & 
						$c_{h_{125} t \bar t}$ & $c_{h_{125} b \bar b}$&$c_{h_{125} VV} $ &$c_{A t \bar t} $ &$c_{A b \bar b} $ \\
						\noalign{\hrule height 0.9pt} 
						{\color{lawngreen}\FiveStar\hspace{-1em}\color{black}\textbf{\FiveStarOpen}}  & 0.41 & -0.31 &   -0.16 &     1.08 &     0.96 &     0.99 & 0.14 & 0.58 \\
						{\color{magenta}\FiveStar\hspace{-1em}\color{black}\textbf{\FiveStarOpen}}  &  0.37 & -0.40 &   -0.19 &     1.09 &     0.94 &     0.98 & 0.21 & 0.57 \\\noalign{\hrule height 0.4pt}
						\noalign{\vspace{1.25pt}}  \noalign{\hrule height 0.9pt}
				\end{tabular*}}
				{	\setlength{\tabcolsep}{0.82cm}\begin{tabular*}{\textwidth}{@{\extracolsep{0.5pt}}lcccccc}
						\textbf{Branching ratios in \%} \\
						$\bm{h}$ & $gg$& $b \bar b$ & $\tau^+ \tau^-$& $\gamma\gamma$ & $W^+W^-$ & $ZZ$  \\
						\noalign{\hrule height 0.9pt}	
					{{\color{lawngreen}\FiveStar\hspace{-1em}\color{black}\textbf{\FiveStarOpen}} }
						& 12.29 &   74.32 &   10.79 &    0.13 &    0.10 &    0.01 \\
						{\color{magenta}\FiveStar\hspace{-1em}\color{black}\textbf{\FiveStarOpen}} & 5.97 &   68.26 &   22.85 &    0.08 &    0.09 &    0.01\\
						\noalign{\hrule height 0.4pt}
						\noalign{\vspace{1.25pt}}  \noalign{\hrule height 0.9pt}
						$\bm{H}$ & $gg$& $b \bar b$ & $\tau^+ \tau^-$ & $\gamma\gamma$ & $W^+W^-$ & $ZZ$  \\
						\noalign{\hrule height 0.9pt}
						{\color{lawngreen}\FiveStar\hspace{-1em}\color{black}\textbf{\FiveStarOpen}} &  8.78 &   58.66 &    6.98 &    0.18 &   19.27 &    2.42 \\
						{\color{magenta}\FiveStar\hspace{-1em}\color{black}\textbf{\FiveStarOpen}} & 9.46 &   59.23 &    4.83 &    0.19 &   20.13 &    2.52\\
						
						\noalign{\hrule height 0.4pt}
						\noalign{\vspace{1.25pt}}  \noalign{\hrule height 0.9pt}
						$\boldsymbol{A}$ &  $gg$& $b \bar b$ & $\tau^+ \tau^-$ & $\gamma\gamma$ \\
						\noalign{\hrule height 0.9pt}
						{\color{lawngreen}\FiveStar\hspace{-1em}\color{black}\textbf{\FiveStarOpen}} &51.80 &      18.97 &   23.72 &    0.10\\
						{\color{magenta}\FiveStar\hspace{-1em}\color{black}\textbf{\FiveStarOpen}} &  41.23 &       33.14 &   21.66 &    0.07 \\
						
						\noalign{\hrule height 0.4pt}
						\noalign{\vspace{1.25pt}}  	\noalign{\hrule height 0.9pt}
						$\boldsymbol{H^\pm}$ & $tb$ &$\tau^\pm\nu$ & $W^\pm  h$ & $W^\pm  H$ & $W^\pm A$ \\
						\noalign{\hrule height 0.9pt}
						{\color{lawngreen}\FiveStar\hspace{-1em}\color{black}\textbf{\FiveStarOpen}}  & 23.92 &    2.27 &   36.69 &    0.05 &   36.58\\
						{\color{magenta}\FiveStar\hspace{-1em}\color{black}\textbf{\FiveStarOpen}} & 23.44 &    2.52 &   34.05 &    0.07 &   39.48 \\\noalign{\hrule height 0.4pt}
						\noalign{\vspace{1.25pt}}  
						\noalign{\hrule height 0.9pt}
						
				\end{tabular*}}	
						{	\setlength{\tabcolsep}{0.82cm}\begin{tabular*}{\textwidth}{@{\extracolsep{0.5pt}}lcccccc}
					\textbf{$B$-physics observables (95\% CL)} \\
					
					\noalign{\hrule height 0.4pt}
					\noalign{\vspace{1.25pt}}  \noalign{\hrule height 0.9pt}
					$\mathcal{BR}$   &  $\bar B\to X_s\gamma$& $B_s\to\mu^+\mu^-$ & $B^0\to\mu^+\mu^-$ & $B^+\to\tau\nu_{\tau}$ \\
					\noalign{\hrule height 0.9pt}
					{\color{lawngreen}\FiveStar\hspace{-1em}\color{black}\textbf{\FiveStarOpen}} &0.000360 &      3.247 $\times 10^{-9}$  &   0.9689$\times 10^{-10}$ &    0.000084\\
					{\color{magenta}\FiveStar\hspace{-1em}\color{black}\textbf{\FiveStarOpen}} &  0.000336 &       3.325$\times 10^{-9}$ &   0.9917$\times 10^{-10}$ &    0.000084 \\\noalign{\hrule height 0.4pt}
					\noalign{\vspace{1.25pt}}  
					\noalign{\hrule height 0.9pt}
					
			\end{tabular*}}	
			\caption{\small The full description of our best fit points.}
			\label{tab:BF}
		}
\end{table*}

\section{Conclusion}\label{sec:con}
Extensive data samples collected by LHC experiments have facilitated detailed analyses of the reported 95 GeV excesses following their initial observations. A rigorous examination of this data, coupled with in-depth simulations and advanced computational techniques, has been conducted. In this context, we have introduced the 2HDM Type-III with a specific Yukawa texture as a theoretical framework for potentially the observed $\gamma\gamma$, $\tau\tau$ and $b\bar b$ anomalies. This model focuses on a Higgs boson with a mass of approximately 95 GeV, produced via gluon-gluon fusion at the 13 TeV LHC and decaying into $\tau\tau$ and $\gamma\gamma$, as well as being produced through Higgs-strahlung at LEP and decaying into $b\bar b$.

Assuming that the heavy CP-even $H$ state in our model is the 125 GeV Higgs boson discovered at the LHC, we have explored parameter spaces where the 95 GeV CP-odd state, $A$, comprehensively explains the LHC excesses at a 1$\sigma$ level, consistent with current theoretical and experimental constraints. We have also demonstrated that the superposition of the light CP-even state $h$ and the CP-odd state $A$ can account for the anomalies observed at both the LHC and LEP, through a $\chi^2$ analysis at the 1$\sigma$ level.

Further analysis confirms that effectively addressing these discrepancies requires an enhancement of the ${t\bar t}H$ coupling, deviating from SM predictions. With upcoming advancements at the HL-LHC, precise measurements are expected to clearly differentiate between the SM-like properties of the $H$ state and the predictions of the 2HDM Type-III. This differentiation is critical for data points showing significant deviations, particularly with the enhanced ${t\bar t}H$ coupling parameter space. Such measurements will be crucial in conclusively confirming or refuting our model. We have provided detailed descriptions of our best fit points to aid further phenomenological studies.

\section{Acknowledgments}
SM is supported in part through the NExT Institute and the STFC Consolidated Grant ST/L000296/1.
We thank A. Belyaev,  M. Chakraborti and S. Semlali for useful discussions.
\newpage
\bibliography{main}
\bibliographystyle{apsrev4-2}
\end{document}